\documentclass[draft]{aipproc}
\layoutstyle{8x11single}

\newcommand{\msun}{M_{\odot}}
\newcommand{\eos}{equation of state~}

\newcommand{\eosp}{equation of state}

\newcommand{\beqn}{\begin{eqnarray}}
\newcommand{\eeqn}{\end{eqnarray}}
\newcommand{\doe}
{This work was supported by the
Director, Office of Energy Research,
Office of High Energy
and Nuclear Physics,
Division of Nuclear Physics,
of the U.S. Department of Energy under Contract
DE-AC03-76SF00098.}

\begin{document}

\author{Norman K. Glendenning}{
  address={Nuclear Science Division, \&
Institute for Nuclear and Particle Astrophysics,\\  
Lawrence Berkeley National Laboratory,
University of California, Berkeley, CA 94720},
  email={},
}

\author{Fridolin Weber}{
  address={Department of Physics, University of Notre Dame,
Notre Dame, IN 46556, USA},
  email={},
  homepage={},
}

\title{Spin Clustering of Accreting X-ray Neutron Stars
as Possible Evidence of Quark Matter}

\date{\today}

\begin{abstract}
  A neutron star in binary orbit with a low-mass non-degenerate companion becomes a source of x-rays with millisecond variability when mass accretion spins it up. Centrifugally driven changes in density profile may initiate a phase transition 
in a growing region of the core  parallel to what  may take place in an isolated millisecond
pulsar, but in reverse. Such a star will spend a longer time in the  spin
frequency range over which the transition occurs
than elsewhere because the change of phase, paced by the spinup rate, is accompanied by a  growth in the moment of inertia.  The
 population of accreters will exhibit a clustering
in the critical  frequency range. A phase change triggered by changing spin and the
accompanying adjustment of moment of inertia has its analogue in 
rotating nuclei.
\end{abstract}

\maketitle




\section{Background}
A neutron star, born with a non-degenerate companion,
initially spins down
because of the drag of the magnetic
dipole torque. A wind from the hot surface
of the 
companion disperses the pulsed  signal and it is radio silent.
Later in 
 the companion's evolution when it overflows the Roche lobe, mass
transfer commences to spin up the neutron star. 
It has begun what is believed to
be 
 its evolution from an
old star with long period
and fairly high magnetic field to a millisecond (ms) pulsar 
with low field
\cite{alpar82:a,heuvel91:a,klis98:b,chakrabarty98:a}. During the intermediate stage it emits x-rays because
the surface and accretion
ring are heated to high temperature.

Accreting canonical neutron  stars would take only $\sim  10^8$ y at an accretion rate of $\dot{M}=10^{-9}  \msun$/y to attain a period of 2 ms (=500 Hz) and any asymmetry in the accretion pattern  would cause millisecond variability in x-ray emission as was foreseen many years ago \cite{Shvartsman71:a,Sunyaev73:a}.  This expectation has been realized in the last several years in numerous discoveries made with the Rossi X-ray Timing Explorer (RXTE). (See van der Klis for a critical  review and discussion \cite{klis00:a}.) One of the big surprises is that most of the  frequencies
observed in these objects, generally interpreted as spin frequencies, seem to be clustered in a small band, or spike at $\sim 300$ Hz.

Several suggestions as to the origin of frequency clustering
and much relevant
research has been published. 
In one scenario, a small quadrupole 
distortion provided by a thermally induced
density asymmetry  creates gravitational waves whose torque
balances that applied by accreting matter
at the observed critical frequencies 
\cite{bildsten98:a}. 
Alternately, Rossby waves may be excited in the crust of the 
neutron star which trigger a thermal runaway of the r-mode, reducing 
the spin below the
excitation frequency, at which time accretion may again
spin up the star. This cycle may be repeated several times before 
the donor star is consumed \cite{anderson2000:a}.
We propose yet another mechanism---a natural extension of 
our earlier
work---that could {\sl temporarily} stall
spinup for an epoch of $10^7{\rm~to~}19^9$ y
 and therefore lead to a frequency clustering in the
population, while
allowing a further evolution to the ms pulsar stage
\cite{glen00:a}.

\section{Role of quark phase}
 The density in the interior of neutron stars is a few times nuclear density. 
At such densities it is quite plausible that  quarks lose their association with particular hadrons---the more 
compressible deconfined quark  matter phase replaces the normal phase  in the core of the star.  
Such a quark matter core does not endow the hybrid  star itself 
with any remarkable property  aside from reducing the limiting  
mass, generally  to values  $\leq 1.5 \msun$---small  compared 
to models of neutron stars that are made purely of  neutrons---but  quite in agreement with observed masses \cite{thorsett99:a}.  Moreover, there  are grounds to believe that 
neutron star masses do  in fact fall  in a very small interval, bounded from below by the  Chandrasekhar limit on the iron core mass  in the pre-supernova star, and above by the    neutron star   mass    which is limited by any 
one of three possible    phase transitions, hyperonization, kaon condensation and    quark deconfinement \cite{glen91:c,glen91:a,brown94:a}. We shall assume therefore that    canonical pulsars---like the Crab and more slowly rotating  ones---have    a quark matter core essentially from birth and that neutron  stars fall    in a narrow mass range.

  By comparison, millisecond pulsars are centrifugally flattened  in the  equatorial plane and the density is diluted in the interior. We shall suppose that the critical phase transition density lies between the  diluted density of ms pulsars and the  density at the center of canonical pulsars. Then as a ms pulsar spins down, or as a canonical neutron star at some  stage begins accreting matter from a companion and is spun up,  a change in density distribution paced by the 
changing centrifugal force will, in some critical
frequency range,
 cause a change
of phase of matter in an expanding region
of the core. In the case of spindown of an  isolated ms pulsar, self-gravity
and the weight of the surrounding part of the  star will  squeeze the more compressible  high-density phase that is forming in the interior.   Conversely, an accreting neutron star that is being spun up,  will, over time,  spin  out the  already present quark phase. In either case, 
the moment of inertia will progressively
alter with the change of phase of matter
and therefore  the star's spin rate will adjust to conserve angular  momentum that is not being carried off by radiation   or supplied by accreted matter fast enough.
 For this reason, a spin anomaly
should occur in both types of objects,
ms pulsars and x-ray neutron stars in binaries,
if it occurs in either. The phase change manifests itself
 as a temporary governer on spin causing changes in rotational 
frequency to stall 
in about the same range in the two types
of objects. 
The   population of x-ray accreters should exhibit more objects 
in the critical frequency range than in neighboring ones. 
This appears to be the meaning of recent discoveries
made with the RXTE  \cite{klis00:a}.
The effect of a phase transition
on the population of ms pulsars is less direct since
it involves a convolution of our results for x-ray accreters. 

Previously, it was found for ms pulsars
that the mixed phase in a model star 
converts to pure quark matter, first at the center and then in an expanding region, paced by the slow loss of angular momentum to radiation \cite{glen97:a,glen97:e,heiselberg98:a,blaschke99:a}.  The consequent decrease in the moment of inertia could even 
introduce an era of {\sl spinup}
 lasting for $\sim 2\times 10^7$ years  or $\sim 1/50$ of the spindown time \cite{glen97:e}. The anomalous spinup of our model star occurred  in a small frequency band around 220 Hz \cite{glen97:a}. 
Such a response of the moment of inertia to a change of 
phase occasioned by changing spin is very like the so-called ``backbending'' in rotating  nuclei
caused by coriolis quenching of BCS nucleon spin pairing 
predicted by Mottelson and Valatin 
\cite{mottelson60:a} 
and 
discovered in the 1970s \cite{johnson72:a,stephens72:a}. (Compare
Figs.\ \ref{nucleusf} and \ref{oif}.)

\begin{ltxfigure}\begin{center}
 \resizebox{.8\columnwidth}{!}
  {\includegraphics{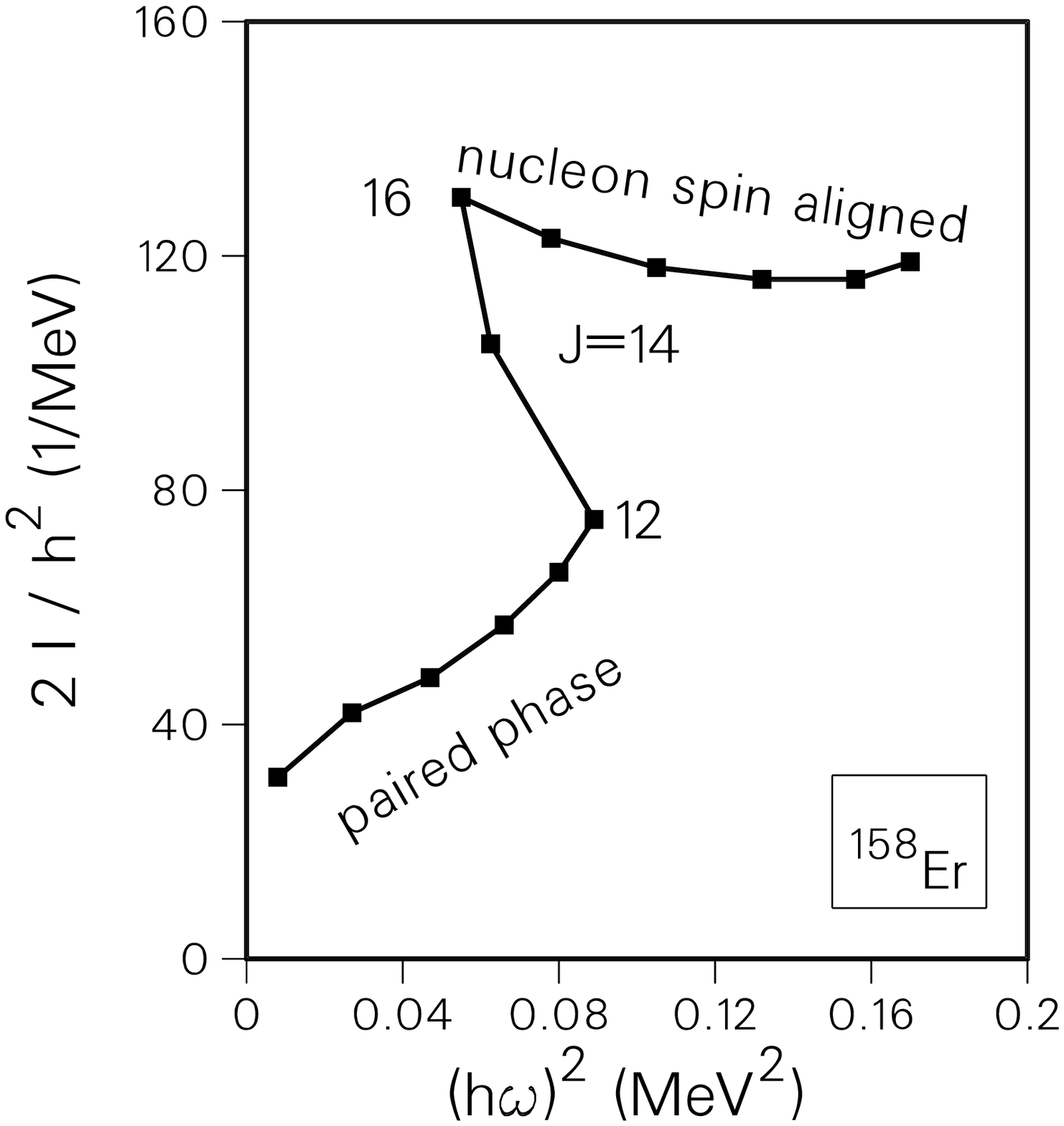}
\hspace{2cm}\includegraphics{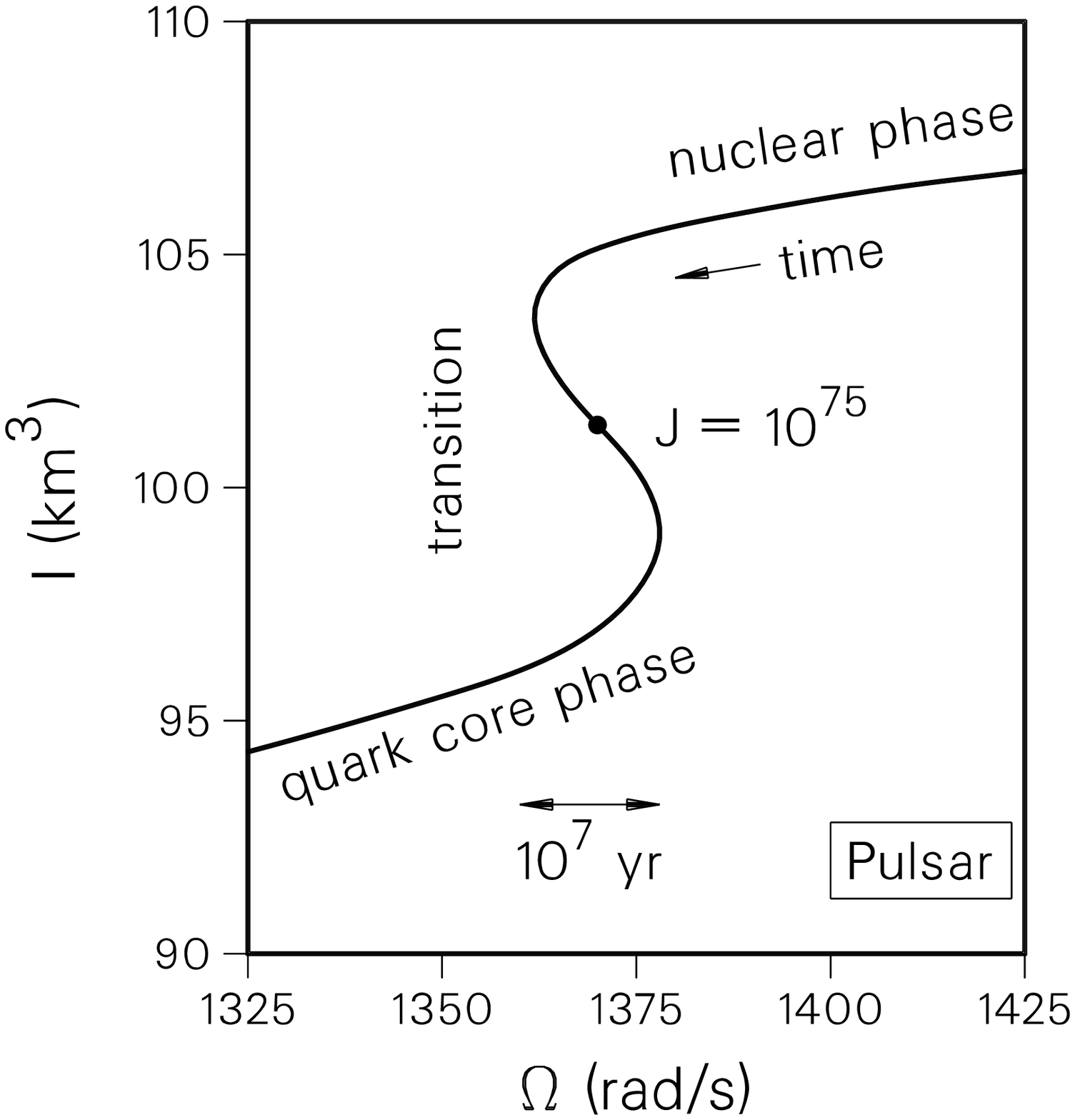} }
{ \caption{ Backbending of moment of inertia in a
radiating  nucleus 
caused by a phase transition between a spin-aligned
and BCS paired state. \label{nucleusf} 
}}
 {\caption{  \label{oif} Backbending in a model neutron star
caused by a phase transition between confined and deconfined
matter. Depending on the stellar mass and \eosp, the transition
may be less singular.
}}
\end{center}
\end{ltxfigure}

\section{Accretion induced spinup}  It is clear from the foregoing  discussion that in our model,  the particular details of the accretion process do not determine in what  range   the spinup stalls for a time. We therefore use a simple schematic model of accretion in which the  spin-up torque of the accreting matter causes a change in the star's angular momentum $J$ according to the relation \cite{elsner77:a,ghosh77:a,lipunov92:book} \begin{eqnarray} {{dJ} \over {d t}} = {\dot M}  \sqrt{M r_{\rm m}} -   \kappa \, \mu^2 \, r_{\rm c}^{-3} \label{eq:dJdt} \end{eqnarray} ($G=c=1$)   ($\kappa\sim 0.1$). The first term represents the torque applied by the accreting matter and the second by the magnetic field of the neutron star and the viscosity of matter in the accretion ring. The star's magnetic moment is denoted by  $\mu \equiv R^3 B$,
the co-rotating
radius by $r_{\rm c} = ( M/ \Omega^{2} )^{1/3}$, the inner edge of the accretion ring by $r_{\rm m} = \xi \, r_{\rm A}$, $(\xi \sim 1)$ 
and the  
Alf\'en radius 
at which the magnetic energy 
density equals the total kinetic energy density of the accreting matter
by $r_{\rm A} = [  {\mu^4} /( 2 M \dot{M}^2) ]^{1/7}$. 

The above equation  can be written as a time evolution equation for the angular velocity  $\Omega$ of the accreting star; 
 \begin{eqnarray} & &  I(t) {{d\Omega(t)} \over {d t}} = {\dot M}  \sqrt{M(t) r_{\rm m}(t)}   - \Omega(t)     {{dI(t)}\over{dt}} - \kappa \, \mu(t)^2 \, r_{\rm c}(t)^{-3} \, .     \label{eq:dOdt.1}\label{spinevolution}    
 \end{eqnarray}     
The moment of inertia $I$ of ms pulsars or of neutron star accreters has to be computed in GR without making the usual assumption of slow rotation.  We use a  previously obtained  expression for the moment of inertia of a rotating star
 \cite{glen92:b}.  

\begin{ltxfigure}[b]\begin{center}
 \resizebox{.7\columnwidth}{!}
  {\includegraphics{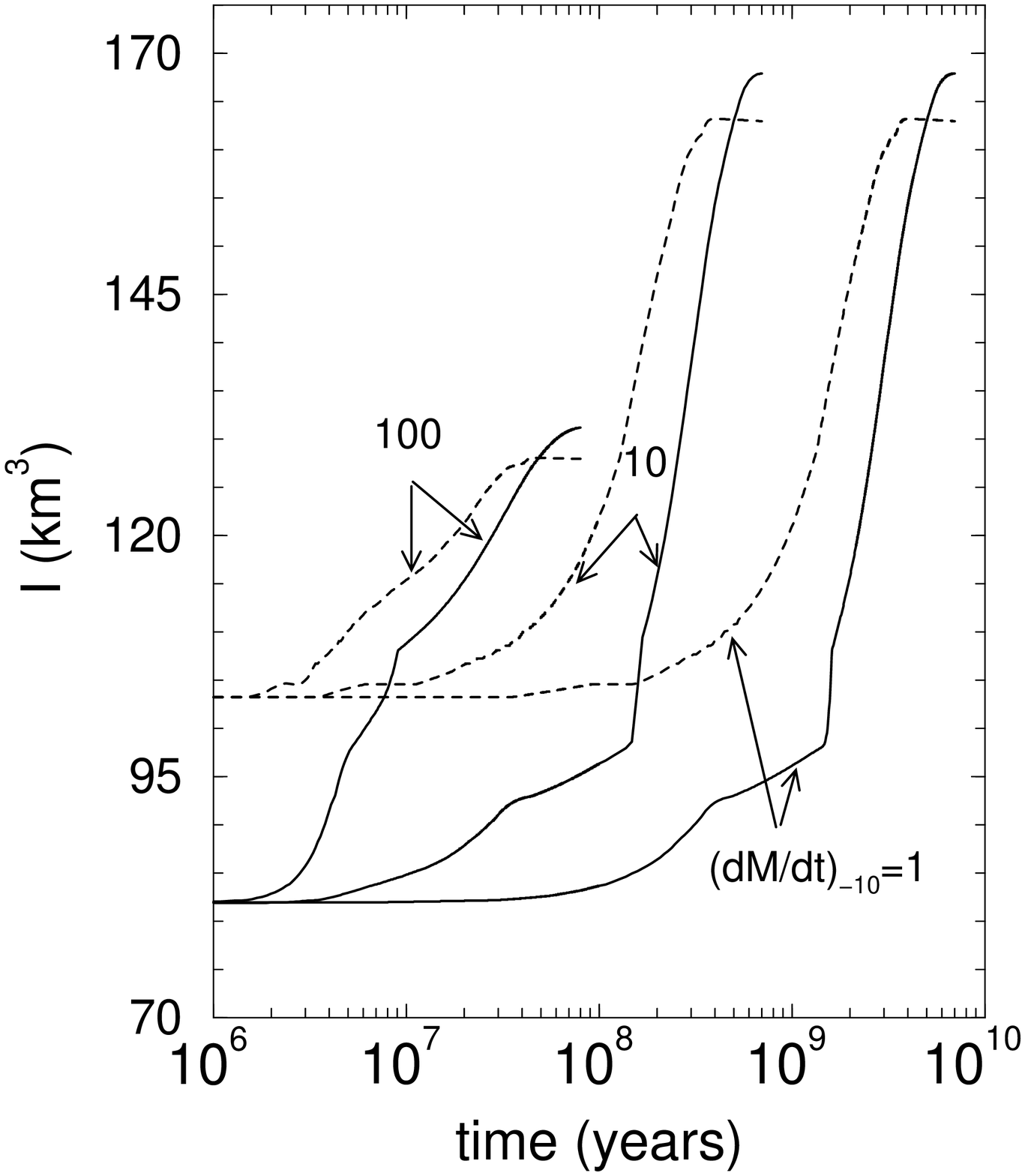}
   \hspace{3cm}          \includegraphics{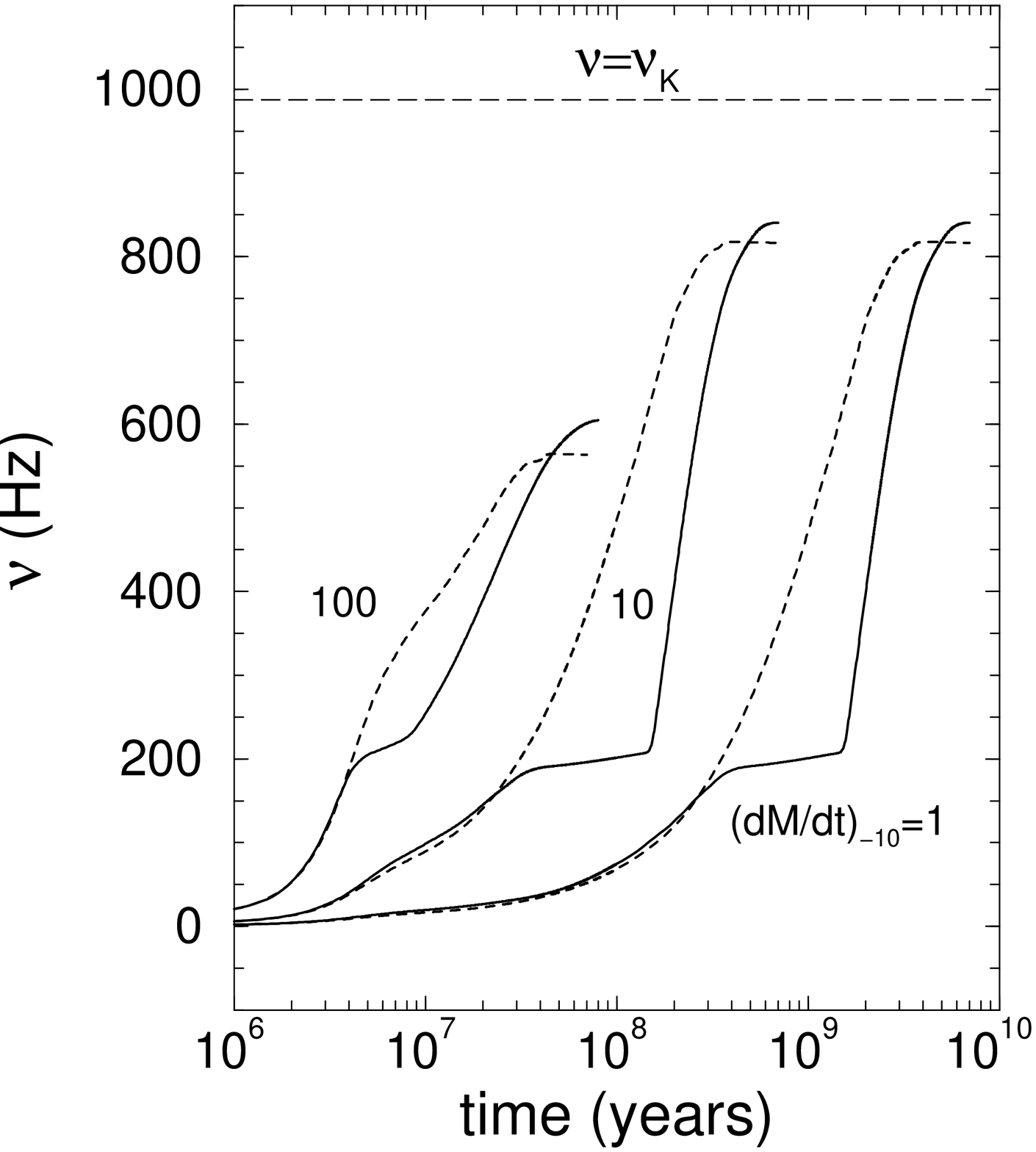} }
{ \caption{  Moment of inertia of neutron stars with  (solid curves)  and without   (dashed curves) quark phase transition   assuming $0.4 M_\odot$ is accreted. Results for three {\sl average} accretion rates are shown. \label{fig:It}  }}
 {\caption{  Evolution of spin frequencies of accreting neutron stars with (solid curves) and without (dashed curves) quark deconfinement if $0.4 M_\odot$ is accreted.  The spin plateau around $200$~Hz signals the ongoing process of quark confinement in the stellar centers.  Spin equilibrium  is eventually  reached. \label{fig:nue} 
}}
\end{center}
\end{ltxfigure}

The  magnetic field $B$ is believed to decay only weakly due to ohmic resistance in canonical pulsars, but very significantly while accreting matter from a companion. This era can last up to $10^9$ y and cause field decay by several orders of magnitude. For a review of the literature and several evolutionary scenarios, see Ref.\ 
 \cite{konar}. Although there is no consensus concerning the magnetic field decay, observationally, we know that canonical pulsars have fields of $\sim 10^{11} {\rm~to~} 10^{13} G$, while ms pulsars have fields that lie in the range $\sim 10^{8} {\rm~to~} 10^{9} G$. We shall rely on this observational fact, and assume that the field decays according to 
$B(t) = B(\infty) + [B(0) - B(\infty)] e^{-t/t_{\rm d}}$
with $t=0$ at the start of accretion, $B(0)=10^{12}$ G,~$B(\infty)= 10^8$ G, and $t_{\rm d} =  10^6$ yr.   Such  a decay to an asymptotic value seems to be a feature of some treatments of the magnetic field evolution \cite{konar}. 
The frequency attained after a few million years of accretion
will be 
independent of the initial value. We take $\nu(0)=1$ Hz. 

The theory  and parameters used to describe our model neutron star
 are precisely those used in previous publications
\cite{glen97:a}. Its initial mass is $M=1.42 \msun$, 
close to the mass limit
of the rotating star of $1.66 \msun$.
Quark matter is treated in a version of the MIT bag model with the three light flavor quarks ($m_u=m_d=0,~m_s=150$ MeV) as described in Ref.\ \cite{farhi84:a}.  A value of the bag constant $B^{1/4}=180$ MeV is employed, as in \cite{glen97:a}. The transition between these two phases of a medium with two  independent conserved charges (baryon and electric) is described in Ref.\ \cite{glen91:a}.

 Figure \ref{fig:It} shows how the moment of inertia changes for  a neutron star in a
binary system that is spun up by mass accretion according  to Eq.\  (\ref{eq:dOdt.1}). In one case we assume that a phase transition between quark matter and confined hadronic matter occurs, and in the other that it does not. This accounts for the different  initial moments of inertia, and also, as we see, the response to spinup. Three {\sl average}
accretion rates are assumed, $\dot{M}_{-10}=1$, 10 and  100 (where $ \dot{M}_{-10}$ is  in units of $10^{-10} \msun$/y).    The corresponding spin evolution of accreting  neutron stars as determined by the changing moment of inertia and the   evolution equation (\ref{spinevolution}) is shown in Fig.\ \ref{fig:nue}.   In both Figs.\ \ref{fig:It} and  \ref{fig:nue}  we assume that $0.4 M_{\odot}$ is accreted. Otherwise the maximum frequency attained is less.

   We compute a frequency distribution  of x-ray stars
in low-mass binaries (LMXBs) from Fig.\ \ref{fig:nue}, for {\sl one} accretion rate, by assuming that neutron stars begin their accretion evolution at the average rate of one per million years.  A different rate will only shift some neutron stars from one bin to an adjacent one.   The donor masses in the binaries are believed to range between  $0.1 {\rm~and~} 0.4 \msun$ and we assume a uniform distribution   in this range.
The resulting  frequency distribution of x-ray neutron stars is shown in Fig.\ \ref{fig:bin}; it  is striking. Spinout of the quark matter core as the neutron star spins up is signalled by a  spike in the  distribution which would be absent if there were no phase transition in our model of the neutron star. The position of the spike depends
only on the stellar model.
 But the weight of the spike as compared to the high frequency tail depends sensitively on the weight with which the donor masses are assigned, the initial mass function of the 
accreting neutron stars (for which we have taken only one mass),
and  to a minor degree
on the accretion rate. A donor of mass $0.1 \msun$ 
contributes only to the
spike, while all greater masses contribute to the spike and
to higher frequency x-ray stars. Objects above about 400 Hz
are unstable to collapse to very high-spin black holes.
Accretors of lower initial mass than we assume would contribute to the long
high-frequency tail as well, possibly, to the spike.
\begin{ltxfigure}\begin{center}
 \resizebox{.8\columnwidth}{!}
  {\includegraphics{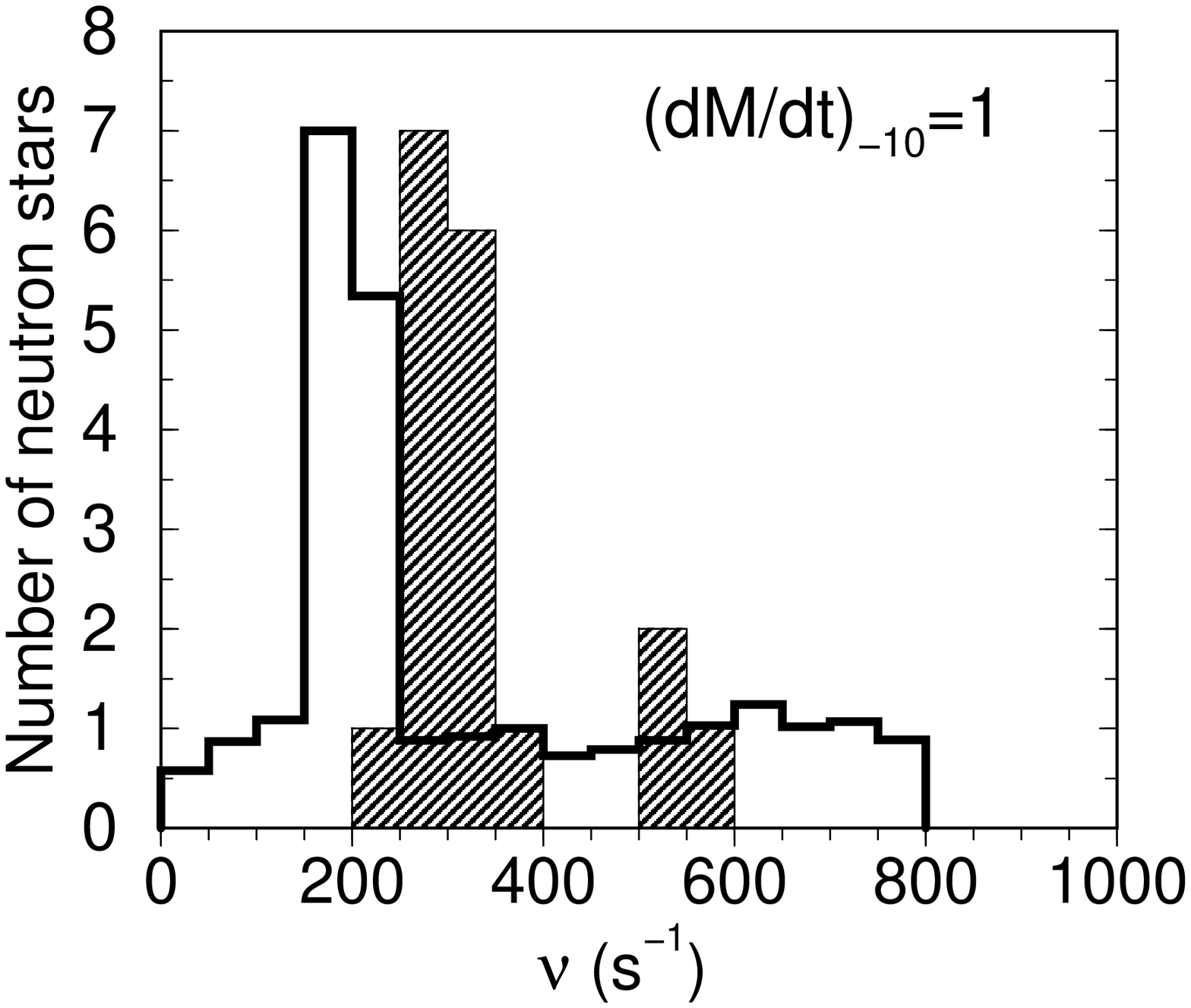}\includegraphics{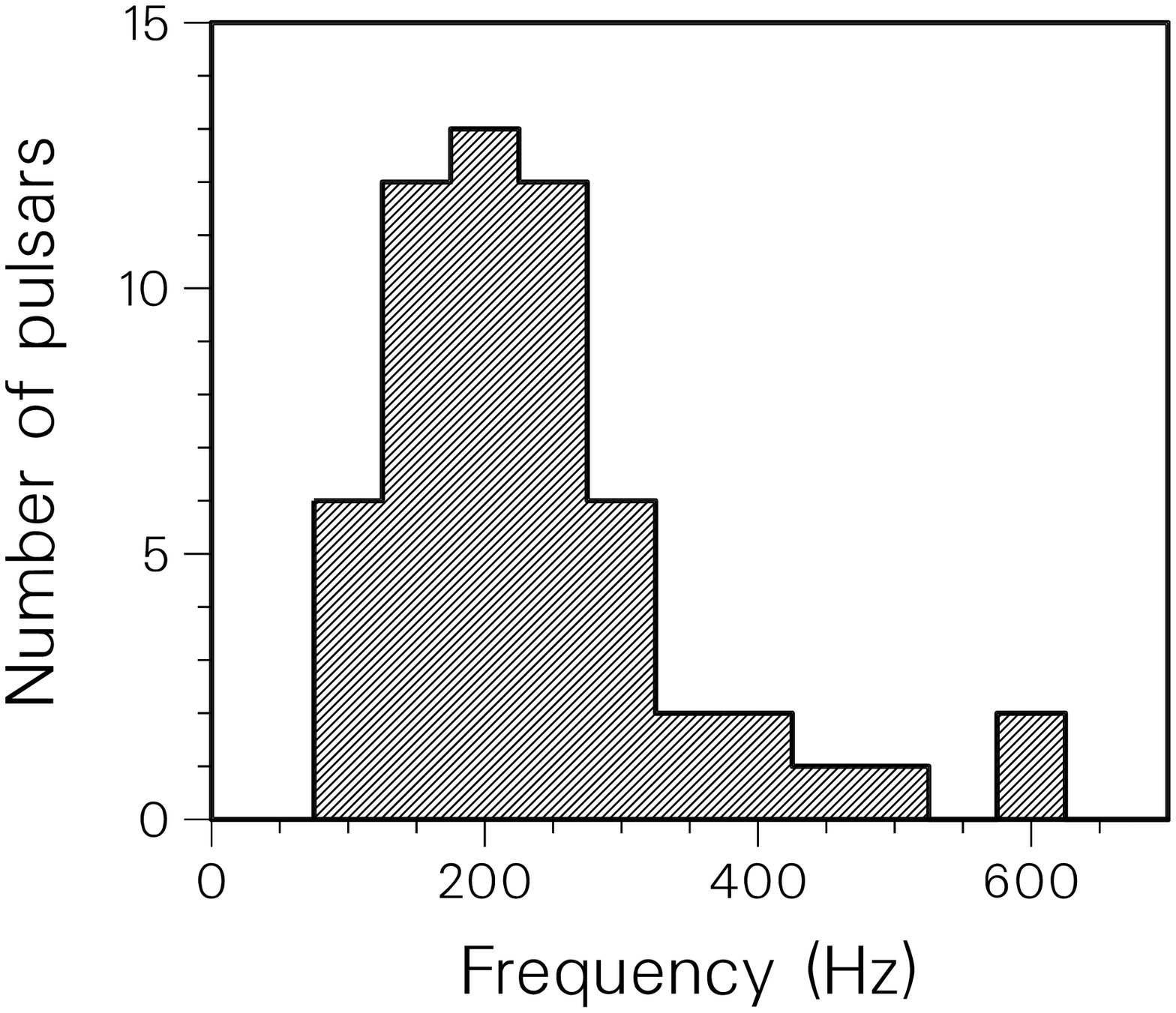} }
{ \caption{Calculated
spin  distribution  of the underlying population  
of x-ray neutron stars for one accretion rate
(open histogram)  is normalized to the number of observed objects (18) at the peak.   Data on 18
 neutron stars in  LMXBs (shaded histogram) is from Ref.\  \protect\cite{klis00:a}.   The spike in the calculated distribution  corresponds to the spinout of the quark matter phase.   Otherwise the spike  would be absent. \label{fig:bin}  
}}
 {\caption{Data on the frequency distribution of 60
millisecond pulsars ($1 \leq P <  10$ ms). The 
frequency bins are 50 Hz wide.  \label{fig:princeton} \label{nu}
}}
\end{center}
\end{ltxfigure}
\section{Discussion and Summary}
Theoretically, a phase transition can (but not necessarily
does) cause a distinct clustering in frequency
of x-ray accreters, which is independent of details of accretion, 
such as rate, mass accreted so long as it exceeds a small minimum
value ($\sim 0.1\msun$),
or indeed to the particular description of 
accretion mechanism that we employ. As emphasized, the position of 
the peak is an intrinsic property of our model star.
 But the transition can also occur 
unheralded by
any remarkable signal \cite{glen97:e}.

The apparent frequency clustering of x-ray neutron stars  is about 100 Hz higher than calculated.
This discrepancy should not be surprising in view of our  ignorance of the \eos above saturation density of nuclear matter and the necessarily crude representation of hadronic matter in the two phases in the absence of relevant solutions to the fundamental QCD theory of strong interactions. But however crude any  model
 of hadronic matter may be, the physics underlying the effect of a phase transition on spin rate is robust, although not inevitable. We have cited an analogous phenomenon  discovered in rotating nuclei
\cite{mottelson60:a,johnson72:a,stephens72:a}.

The data in Fig.\ \ref{fig:bin} is gathered from Tables 2--4 of the review article of van der Klis concerning discoveries made with  the  Rossi X-ray Timing Explorer \cite{klis00:a}. The interpretation of millisecond oscillations in the x-ray emission, either that  found in  bursts or of the difference between twin  quasi-periodic oscillations in x-ray brightness, is  ambiguous in some cases.  For example, 
some of
the burst
data near 600 Hz may actually represent twice the rotational frequency of the star. For this and other caveats, see the
review article \cite{klis00:a}.

Nevertheless, the basic feature will probably survive---a clustering of x-ray neutron stars at moderate spin and a high spin tail. Certainly there are high spin {\sl pulsars}.   A histogram of {\sl ms pulsar} frequencies shows  a concentration 
around  200 Hz, and a  tail extending to $\sim 600$ Hz  as shown in Fig.\ \ref{fig:princeton}. So both the (sparse) data on x-ray objects and on ms pulsars seem to  agree on a peak  in the number of stars at moderate spin
and on   attenuation at  high spin.  For ms pulsars the attenuation may be   partly a selection effect due to interstellar dispersion of the radio signal.

 To summarize, we suggest that the apparent clustering in rotation frequency of accreting x-ray neutron stars in low-mass binaries may be
 caused by  the progressive conversion of quark matter in the core to confined hadronic matter, paced by the slow spinup
due to mass accretion. 
When conversion is completed, normal accretion driven spinup resumes.
 To distinguish this conjecture from others, one would have to 
discover the inverse phenomenon---a spin anomaly near the
same frequency
in an isolated ms pulsar \cite{glen97:a}.  
If such a discovery were made, and
the apparent clustering of x-ray accreters is confirmed, we would
have some degree of confidence in the hypothesis
that a phase of matter such as
existed in the very early universe, is reformed in a cold
state during the birth of
neutron stars. \\

 \doe \vspace{-.2in}


\end{document}